\documentclass{llncs}
\usepackage[T1]{fontenc}
\usepackage{latexsym}
\usepackage{xspace}
\usepackage{wrapfig}
\usepackage{mathptmx}
\usepackage{theorem}
\usepackage{multicol}
\usepackage{amsmath}
\usepackage{tikz}
\usepackage{tikz-qtree}
\usepackage{url}

\newcommand\assign{\leftarrow}
\newcommand{\round}[1]{[#1]}

\newcommand\rss{\textsc{rsss}\xspace}

\newlength\leafspc
\newcommand\emitnode[2]{\rlap{\raisebox{-11pt}[0pt][9pt]{\hspace{-.25em}\scriptsize \hbox to 2em {\hfil #2\hfil}}}\framebox{#1}\,}
\newcommand\leafelt[1]{\rotatebox{270}{\scriptsize $#1$}}

\title{Integer Set Compression and Statistical Modeling}
\author{N.\,Jesper Larsson}

\institute{IT University of Copenhagen, Denmark,\\
\email{jesl@itu.dk}}

\date{}
\begin{document}
\maketitle

\begin{abstract}\let\smaller\footnotesize
  Compression of integer sets and sequences has been extensively studied for settings where elements follow a uniform probability distribution. In addition, methods exist that exploit clustering of elements in order to achieve higher compression performance. In this work, we address the case where enumeration of elements may be arbitrary or random, but where statistics is kept in order to estimate probabilities of elements. We present a recursive subset-size encoding method that is able to benefit from statistics, explore the effects of permuting the enumeration order based on element probabilities, and discuss general properties and possibilities for this class of compression problem.
\end{abstract}

\section{Introduction}\label{sec-intro}

Data compression in its most basic form is commonly expressed in terms of representing a string of characters drawn from a fixed alphabet. A situation with somewhat different characteristics is when the data to represent is a \emph{set}, i.e., a sequence of non-repeating elements whose order is insignificant. Although it is possible to transform the one scenario to the other and vice versa, probability distributions and applicable modeling schemes differ, and there is benefit in treating the problems separately.

This work focuses on compression of sets whose elements are drawn from a fixed range of integers, which we refer to as the \emph{universe}. Another interpretation, common in related work~\cite{MoffatStuiverIntSeqSets,TeuholaTournament,TeuholaInterpol2010} is to view the items to be compressed as the differences between consecutive elements in sorted order, i.e., a sequence of integers. The same encoding methods can be described in terms of set or integer sequence compression~\cite{MoffatStuiverIntSeqSets}. In this work, we prefer the set interpretation, since we are interested in using statistics for individual elements of the universe (as opposed to the gaps between them).

Compression of sets has a number of uses as a component of other compression or data structure problems. One of the more prominent ones is storage of \emph{inverted indexes}~\cite{ZobelMoffatInvertedFiles2006,wmb99:mg}. Others appear in a wide variety of applications such as succinct data structures~\cite{NavarroSuccinct}, data mining~\cite{SiebesVreekenVanLeeuwen}, and web graph representation~\cite{BoldiVignaWWWCodes}. Foundations for coding of sets go many decades back~\cite{EliasCoding,GolombCoding,CoverEnum} and developments stretch into recent work. Of particular interest as related to this work are interpolative coding and related methods~\cite{MoffatStuiverInterp2000,TeuholaInterpol2010,TeuholaTournament} and methods that use binary tries for compressing sets and multisets~\cite{ReznikDST,GriponTries}. There are, however, different classes of modeling assumptions, and works are not generally applicable to the same settings and applications. In particular, little work has been published that attempts to make use of statistics over elements, which is among our main focal points.

This work is outlined as follows. Section~\ref{sec-oldmethods} relates previous methods of particular importance to our work. We note that a slight optimization of cap coding is possible for the fixed-universe set compression. Section~\ref{sec-statistical} presents our method of \emph{recursive subset-size}, relates it to other methods, and discusses statistical set compression issues in general. Section~\ref{sec-concl} concludes and points to future research. Parts of this work have been previously presented in poster form~\cite{larssondccsetsposter}.

\paragraph{Formal Problem and Notation}

In general, data compression can be expressed as encoding a message into a compact format by which it can subsequently be reconstructed by a decoder, using a set of code-specific premises shared by encoder and decoder. We view the encoding process as a sequence of \emph{emit} operations, which each specify an event corresponding to a property of the message. An emit contributes a number of bits to the encoded output. Ideally, emitting an event that has probability $p$ should take $-\log_2 p$~bits~\cite{ShanWeav}. Given that probability ranges of the possible events to be emitted can be inferred in the same way by encoder and decoder, we can use arithmetic coding~\cite{Rub79} to produce a number of bits arbitrarily close to the ideal, even when the desired number of bits is a fractional number or less than one. Hence, we generally assume that ability to estimate probabilities is enough to uniquely define both encoding and decoding.

The special case of encoding an integer $x$ such that $L\le x\le H$ for integers $L$ and $H$, we denote as emitting $x[L, H]$. The bits thus produced depend on the encoding used, and may also depend on probability estimates for the numbers $L,\ldots,H$ shared by encoder and decoder. When $L=H$, zero bits are produced.

We study the problem of encoding a set $S$ consisting of $|S|$ integers drawn from universe $U=0,1,\ldots,|U|-1$. An equivalent interpretation is to view elements as bitstrings whose lengths are limited by $\lceil \log_2 |U|\rceil$.$0,1,\ldots,|U|-1$. We use these interpretations interchangeably. We assume that knowledge of $|U|$, which completely defines $U$, is shared by encoder and decoder. Although we do not generally consider $|S|$ to be known in advance, we do not devote much effort to the encoding of $|S|$. Most of the methods we consider (the only exception being the yes/no code in section~\ref{sec-yesno}) depend on $|S|$ being encoded separately, and its choice of code is independent of the main coding scheme. Section~\ref{sec-somethingaboutssz} does, however, address encoding of $|S|$.

\paragraph{Note on Experiments}

This work is not directed at any particular application area. In order to evaluate performance, we test on primarily three instances of natural data, with different characteristics consisting of small and moderate-sized sets, as well as on some extreme generated data. The first natural data instance, \emph{txt}, tests performance on a very small universe. It takes elements as bits of either individual characters ($|U|=8$) or three bytes grouped together ($|U|=24$). The other two sets are generated from a set of Unix documentation files. In one, \emph{words}, the sets are files, and elements are randomly assigned numbers of the words contained in the set. In the other, \emph{inverted}, each set corresponds to the numbers of the files (randomly assigned) in which the word appears. For \emph{words}, $|U|=19515$ and average $|S|$ is $634$. In \emph{inverted}, $|U|=337$ and average $|S|$ is $11$.

\section{Gap and Range-Narrowing Codes}\label{sec-oldmethods}

This section describes previous methods of particular relevance to our work. Gap coding is the classic methods for independent elements. Range-narrowing methods recursively encode elements, and perform particularly well for clustered elements.

\subsection{Gap Codes}\label{sec-gap}

Set representation can be transformed to sequence representation by arranging the
elements of $S$ in increasing order, and representing a sequence of gaps between adjacent elements.
This is a common technique, described comprehensively e.g. by Witten, Moffat, and Bell~\cite{wmb99:mg}.

Assuming that $|S|$ is encoded separately before the elements, and that all elements are equally likely, we have, for a specific $S$ and any $x\in U$, a global probability $p = \Pr(x\in S) = |S|/|U|$. Hence, the probability of gap size $k$ can be estimated by the geometric distribution~\cite{JaynesProb} as $(1-p)^{k-1}p$. Computing probability ranges in accordance with this distribution, we can achieve minimal encoding length with arithmetic coding~\cite{Rub79}, or a Golomb code~\cite{GolombCoding} that approaches the same property.

We note, however, that geometric distribution is an approximation, corresponding
to draws \emph{with replacement} from a set of size $|U|$ with $|S|$ success
states. In actuality, since elements in a set are distinct, the draws are
\emph{without replacement}. Taking this into account yields a slightly better
estimate. Let $V\subseteq U$ be the part of $U$ that remains after encoding $|S|-n$
elements. Then the probability of the next gap size being $k$ is
$\prod_{i=0}^{k-1}\left(1-n/(|V|-i)\right)\cdot n/(|V|-k)$.
For small $|U|/|S|$ this can yield a noticeable difference, as seen can be seen on the \emph{txt} data results in table~\ref{tab-nonstat}. A similar argument can be used for modifying Golomb~\cite{GolombCoding} or Elias codes~\cite{EliasCoding} to reflect that numbers are chosen from a limited, decreasing, range.

\subsection{Range-Narrowing Codes}\label{sec-rangenarrowing}

Interpolative coding~\cite{MoffatStuiverInterp2000,TeuholaInterpol2010} uses a \emph{low-short} binary
code~\cite{TeuholaTournament} to encode first the highest-numbered element, and then the median element
of $S$. It then progresses recursively in the subsets below and above the
median, always encoding the median, as deeply as necessary to uniquely represent every
element. The size of the set is represented separately.

In terms of compression ratio, the strength of interpolative coding is that if
the elements of $S$ are clustered (i.e. have numbers close together), recursive
progression quickly narrows in on small subsets of $U$, requiring only a few bits for
each binary code.

The closely related \emph{tournament coding}~\cite{TeuholaTournament} is formulated as compression of an integer sequence, corresponding to the differences between consecutive set elements in sorted order. It progresses recursively over the sequence, encoding in each step the maximum element in the range. The original version of tournament coding works for unlimited-size integers,and the global maximum is submitted using Elias' gamma code~\cite{EliasCoding}. In our range-limited setting within a known $|U|$, the maximum is better encoded using the same high-short binary code as the rest of the elements. In our tests, the results are roughly similar to those of interpolative coding, over which Teuhola demonstrates an advantage for uniform distributions.

\section{Recursive Subset-Size Code and Use of Statistics}\label{sec-statistical}

Gap-oriented methods adapt only to the \emph{global} density of elements, based on a single set-size parameter. Range-narrowing methods are able to exploit \emph{local} density differences, by reducing the codeword length for elements. But neither of the methods presents a natural way of exploiting statistical data about the frequencies of individual elements. We now consider coding schemes that do, to varying degrees.

\subsection{Prelude: Yes/No Code and Exponential Statistics}\label{sec-yesno}

Assume that we can predict the probability for the inclusion of every possible element being included in the set to encode, i.e., for every $x\in U$ we have an estimate of $\Pr(x\in S)$. Then arithmetic coding lets us emit $|U|$ \emph{included} or \emph{not included} events, one for each possible element, using the corresponding probability range, by which we obtain a total encoded length of the optimal $-\sum_{x\in U} \Pr(x\in S)$.

Although this is optimal if inclusion in the set is independent among the possible elements, it ignores any correlation between elements. For example, say that two elements $x$ and $y$ usually appear together, i.e., $\Pr(x\in S \land y\in S) > \Pr(x\in S)\times\Pr(y\in S)$. We could address this by keeping statistics on element probability conditioned on inclusion or exclusion of the previous elements in the yes/no encoding order. (By basic laws of conditional probability~\cite{JaynesProb}, the order has no impact on the overall probability estimate of a specific set, and hence neither on the optimal encoding length.) However, this would require statistics whose storage space is exponential in $|U|$, and is hence only realistic for small universes.
\label{sec-unrealisticstats}

In principle, since the yes/no code is neutral as for how $\Pr(x\in S)$ is estimated, it can be used for generating an optimal encoding length given any probability model. However, the neutrality also implies a lack of support for efficiently implementing any particular model. Furthermore, it always requires $|U|$ emissions, which is not efficient for small sets drawn from a large universe. Table~\ref{tab-nonstat} includes encoding lengths for the yes/no code with globally calculated probability estimates as a baseline comparison for the other methods.

The yes/no code can be expected to produce the same encoding length as a Huffman code over the $2^{|U|}$ possible sets, where the probability of a specific set $\tilde{S}$ is $\prod_{x\in \tilde{S}} \Pr(x\in S) \times \prod_{x\not\in \tilde{S}}1-\Pr(x\in S)$. Again, this is a construct exponential in $|U|$, and hence only viable for small universes.

\subsection{Recursive Subset-Size Code}\label{sec-basicrss}

We now present a code that recursively emits subset sizes over the left and right half of the element range, which we refer to as \rss. This is somewhat similar to the range-narrowing codes, but allows the use of individual probability estimates, including a certain degree of context information. The number of counters for maintaining statistical information is bounded by $|U|$, a compromise with feasible space requirements even for large universes.

We begin by describing the basic method without use of statistics.

\begin{figure}[tbp]
\begin{center}
\begin{tikzpicture}
  \tikzset{style=thick}
  \tikzset{level 1/.style={level distance=46pt}}
  \tikzset{level 2/.style={level distance=42pt}}
  \tikzset{level 3/.style={level distance=36pt}}
  \tikzset{level 4/.style={level distance=36pt}}
  \tikzset{level 5/.style={level distance=15pt}}
  \footnotesize
  \Tree
  [ . \emitnode{6}{1}/11
    [ .\emitnode{5}{2}/8
      [ .\emitnode{2}{3}/4
        \edge[style=dotted,thick] ; [ .\emitnode{0}{4}/2
          \edge[style=dotted,thick] ; [ .0 ]
          \edge[style=dotted,thick] ; [ .0 ]]
        [ .2/2
          [ .\emitnode{1}{5} \edge[draw=none] ; [ . \leafelt{0010} ]]
          [ .1 \edge[draw=none] ; [ . \leafelt{0011} ]]]]
      [ .3/4 
        [ .\emitnode{1}{6}/2
          \edge[style=dotted,thick] ; [ .\emitnode{0}{7} ]
          [ .1 \edge[draw=none] ; [ . \leafelt{0101} ]]]
        [ .2/2
          [ . \emitnode{1}{8} \edge[draw=none] ; [ . \leafelt{0110} ]]
          [ . 1 \edge[draw=none] ; [ . \leafelt{0111} ]]]]]
    [ .1/3
      [ .\emitnode{1}{9}/3
        \edge[style=dotted,thick] ; [ .\emitnode{0}{10}/2
          \edge[style=dotted,thick] ; [ . 0 ]
          \edge[style=dotted,thick] ; [ . 0 ]]
        [ .1
          [ . \emitnode{1}{11} \edge[draw=none] ; [ . \leafelt{1010} ]]
          \edge[draw=none] ; [ . \node[draw=none] { } ; ]]]
      \edge[draw=none] ; [ . \node[draw=none] { } ; ]]]
\end{tikzpicture}
\end{center}
\caption{\small\label{fig-rsstree}Binary tree illustrating \rss coding of $S=\{0010,0011,0101,0110,0111,1010\}$ from universe $U=\{0000,\ldots,1010\}$, i.e., $|U|=11$.}
\end{figure}
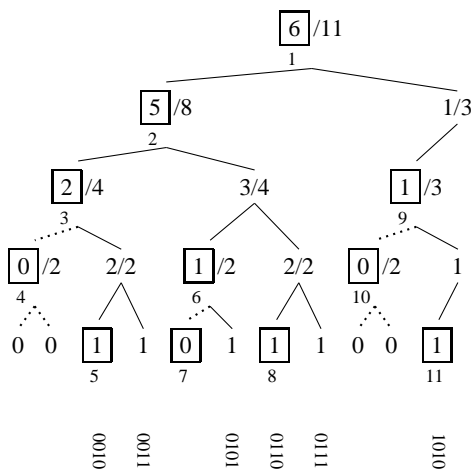

Consider the binary tree over $U$ where $|U|=11$, as shown in figure~\ref{fig-rsstree}. Structurally, this is the complete binary tree with $2^{\lceil \log_2 |U|\rceil} = 16$ leaves, cut off on the right side along the path between the root and the leaf corresponding to $|U|-1$. It can be viewed as an uncompressed binary decision diagram~\cite{LeeBDD}, where each level of the tree is a decision based on one bit of an element, ordered from most to least significant. Each internal node $t$ in the figure is labeled $n_t$/$V_t$, where $V_t$ is the size of the subuniverse (the number of leaves) in the subtree rooted at $t$, and $n_t$ is the size of the subset of $S$ that falls in that subuniverse. Leaves, whose subuniverse size is always $1$, are labeled only with subset size ($0$ or $1$).  We are concerned with representing only the part of the tree corresponding to nonempty subsets (solid-line edges in the figure), which for a sparse set is a relatively small part of the full tree. It can be viewed as a binary trie representing the elements of $S$ (one can note, also the correspondence to the trie-oriented \textsc{dst} code~\cite{ReznikDST}). In another interpretation, it resembles a wavelet tree~\cite{wavelettree}, representing a string of unique symbols in increasing order.

The $V_t$ values depend only on $|U|$. For the root, we have $V_{\mathit{root}}=|U|$. Let $p$ be the root of a subtree of height $h > 0$. The subuniverse sizes of its left and right child are $V_\ell=\min\{2^{h-1}, V_p\}$ and $V_r=V_p-V_\ell$.

We encode the tree by emitting $|S| = n_{\mathit{root}}$ followed by, in a specific top-down order, subset size $n_t$ for each node $t$ that is a \emph{left} child of some $p$ with $n_p>0$. The right sibling of $t$ has subset size $n_p-n_t$, and hence does not need to be explicitly represented.

Traversing the tree top-down recursively narrows the ranges of subset sizes, similarly to the range-narrowing methods of section~\ref{sec-rangenarrowing}. We choose inorder (depth-first, left-to-right) traversal, although any well-defined top-down order would do. The values emitted in the figure example are those shown in frames, and the emit order is shown as numbers below the frames. As for their ranges, we clearly have $0\le n_t\le n_p$, where $p$ is the parent of $t$, but the range can often be bounded further. Let $t$ be a node whose $n_t$ is to be emitted and $r$ its right sibling. Since $n_r\le V_r$ and $n_t=n_p-n_r$, we have $n_t \ge n_p-V_r = n_p + V_t - V_p$. Also, obviously $n_t\le V_t$. Hence, the range of possible values of $n_t$ is $\left[\max\{0, n_p + V_t - V_p\}, \min\{n_p, V_t\}\right]$.

Encoding the example in the figure begins with $n_{\mathit{root}}=6$ in the range $[0,11]$. It then progresses with the left child of the root whose possible subset sizes, by the given computation, is between $3$ and $6$ (inclusive). Hence, the emitted value is $5[3,6]$, and the rest of the sequence is $2[1,4]$, $0[0,2]$, $1[1,1]$, $1[1,2]$, $0[0,1]$, $1[1,1]$, $1[1,1]$, $0[0,1]$, and $1[1,1]$. For completeness, we include emitting in unit-size ranges (such as $[1,1]$) in the sequence, although it produces no bits in the encoding. Decoding works by tracing the same traversal as encoding, and can decode the emitted subset sizes thanks to the top-down order.

The simplest way to encode the $n_t$ would be by a binary code, which makes the code somewhat similar to interpolative coding. However, a flat binary code corresponds to a an implicit assumption of uniform distribution among the possible subset sizes, which would be a peculiar distribution to appear in practice. Hence, compression performance (row~6 in table~\ref{tab-nonstat}) is not difficult to beat.

\paragraph{Uniform Distribution}

A more likely scenario would be uniform distribution among the \emph{elements} in the subset. Let $t$ be a non-root node whose subset size is to be encoded, and $p$ its parent. Let $s=V_t$ and $f=V_p-V_t$. The probability that $n_t=m$ is that of $m$ successes in $n_p$ draws, \emph{without} replacement, from a population of size $V_p=s+f$, whereof $s$ individuals correspond to success and $f$ to failure. This corresponds to \emph{hypergeometric distribution}~\cite{JaynesProb}. We have $\Pr(n_t=m)=\binom{s}{m}\binom{f}{n_p-m}/\binom{s+f}{n_p}$,
%
%
and the expected value of $n_t$ is $n\times s/(s+f)$. We can let this distribution decide probability ranges for arithmetic coding. Note that the distribution does not depend on the individual uniform probability of the elements.

Returning to the example in figure~\ref{fig-rsstree}, let $t$ be the root's left child, whose subset size is~$5$. We have $\Pr(n_t=5))=\binom{8}{5}\binom{3}{6-5}/\binom{8+3}{6}\approx 0.36$. Summing over the possible range of set sizes in this case yields $\sum_{m\in[3,6]} \Pr(n_t=m) = 1$, as we would expect.

In computing probability ranges for $n_{\mathit{root}}=|S|$, the hypergeometric distribution is not useful, since there are no known $s$ and $f\!$. Instead, we could ideally assume that the probability for $S$ having $m$ elements is that of $m$ successes in $|U|$ draws, where the success probability is $p=\Pr(x\in S|x\in U)$. This corresponds to the binomial distribution~\cite{JaynesProb}: $\Pr(|S|=m)=\binom{|U|}{m}p^{m} (1-p)^{|U|-m}$. It is to be expected that for a uniform distribution where the element probability $p$ is known, encoding $S$ using the binomial distribution for $|S|$ and the hypergeometric distribution for each subsequent $n_t$ yields the same total encoding length as the corresponding yes/no code, i.e., $-|S|\log_2 p - (|U|-|S|)\log_2(1-p)$~bits. In particular, setting $p$ to $1/2$ should encode any set in $|U|$~bits.

\label{sec-somethingaboutssz}
In the general case where $p$ is not known, we must resort to a cruder estimate. When the specific application contributes no additional information, assuming uniform distribution over set sizes is perhaps the most reasonable compromise, since it limits the penalty to $\log_2 |U|$ bits per sets. Unless $S$ is very small, this adds relatively little to the encoding size.

Table~\ref{tab-nonstat} shows test results for the methods discussed up to this point. Encoding length is given excluding the representation of $|S|$, except for the yes/no code, which does not require that $|S|$ is represented separately.

Note the similarity  in performance patterns of \rss with the range-narrowing codes, but where \rss appears to have a consistent overhead. We defer a formal analysis to future work, but conjecture that the overhead is at least partly due to how \rss often needs to explicitly encode zero-size subsets, which simply do not appear in the range-narrowing codes. Note also that none of the methods are clearly better than gap coding for this case. We have no indication of \rss contributing to compression performance unless the uniform-probability assumption is surpassed by the use of statistics, which is what we turn to next.

\begin{table}[tbp]
\footnotesize
\begin{center}
\tabcolsep 2pt
\def\arraystretch{1.1}
\begin{tabular*}{.5\textwidth}{@{\extracolsep{\fill}}rlcccc}

&\multicolumn{1}{c}{}& \multicolumn{1}{c}{\textit{txt/8}} & \multicolumn{1}{c}{\textit{txt/24}} &
\multicolumn{1}{c}{\textit{words}} & \multicolumn{1}{c}{\textit{inverted}} \\

\hline
{\scriptsize 1} &\textit{gap} & 1.71 & 2.04 & 5.02 & 4.64 \\
{\scriptsize 2} &\textit{gap w/o repl.} & 1.53 & 1.96 & 5.02 & 4.55 \\
{\scriptsize 3} &\textit{yes/no} & 1.70 & 1.70 & 5.09 & 5.25 \\
{\scriptsize 4} &\textit{interpolative} & 1.65 & 2.16 & 5.43 & 4.82 \\
{\scriptsize 5} &\textit{tournament} & 2.03 & 2.55 & 5.37 & 5.05 \\
{\scriptsize 6} &\textit{\rss{} flat} & 1.99 & 2.61 & 5.60 & 5.12 \\
{\scriptsize 7} &\textit{\rss{} hypergeom.} & 1.53 & 1.96 & 5.02 & 4.55
\end{tabular*}
\end{center}
\caption{\small\label{tab-nonstat}Encoding lengths (bits per element) for non-statistical codes.}
\end{table}

\subsection{Using Statistics}\label{sec-reccstats}

As an example of a statistical scenario, assume that we can store observations taken from a large set of typical sample sets $D=\{S_1,\ldots S_{|D|}\}$. As observed in section~\ref{sec-unrealisticstats}, maintaining counters for producing an individual estimate for each of the possible $2^{|U|}$ sets in unrealistic. Instead, our approach is be to maintain counters in the binary tree over $U$ described in section~\ref{sec-basicrss}. For each left child $t$ in the tree, we maintain $C_t=\sum_i n_{t, i}$, where $n_{t,i}$ denotes the value of $n_t$ (defined in section~\ref{sec-basicrss}) in processing sample set $S_i$. We obtain $q_t=C_t/C_p$, which, on encoding the $n_t$ of a specific set, lets us estimate the expected value of $n_t$ by $q_t n_p$.

Henceforth, we simply assume that correct $q_t$ estimates are available, by some prior knowledge about the distribution of sets. For our performance measurements, we obtain $q_t$ values by counting element appearances, across the set of test inputs for respective tests. This is not intended as a suggestion for practical use, but consider it a choice for testing the capability of our method to adopt probability ranges in accordance with known statistics.

We now consider the use of $q_t$ to find a better estimate for $\Pr(n_t=m)$ for each $m\in \left[\max\{0, n_p - f\}, \min\{n_p, s\}\right]$ (where $s$ and $f$ defined as in section~\ref{sec-basicrss}). It may seem reasonable to use the same hypergeometric distribution as for the uniform assumption, after replacing $s$ and $f$ with $s'=\round{q_t (s+f)}$ and $f' = (s+f)-s'$. However, this would assign zero probability to some possible $n_t$. For instance, let $t$ be the left child of the root in figure~\ref{fig-rsstree}, and say that statistics tells us that $q_t = 0.35$, which yields $s'=\round{0.35\times 11}=4$ and $f'=11-4=7$, and the probability range for the desired value $n_t=5$ is zero.

Instead, we consider the following options.

\paragraph{Binomial Approximation and Case Exclusion}

For large $|U|$, hypergeometric distribution (\emph{without replacement})
approaches the corresponding distribution \emph{with replacement}, i.e. binomial
distribution. Using binomial distribution as an estimate makes it simple to
incorporate $q_t$, by setting $\Pr(n_t=m)=\binom{n_p}{m}q_t^m (1-q_t)^{n_p-m}$. The
desired expected value of $q_tn$ is retained, and all possible values are given nonzero probabilities.

However, this estimate 
assigns nonzero probability to all $0\leq m\leq n_p$, which may include values smaller than
$n_p-f$ and larger than $s$. This is clearly wasteful. For example, for $s=5, f=5, n_p=7$,
more than 80\% of the range is taken by the impossible cases $m=0, m=1$. We adopt the following strategy to adjust
the values of $m,n_p,s,f$ before encoding, to remove the correct number of unusable
states:
\begin{enumerate}
\item If $n_p>s$, reassign, in order, $d\assign n_p-s$, $n_p\assign s$, and $f\assign f-d$.
\item Then, if $n_p>f$, reassign, in order, $d\assign n_p-f$, $m\assign m-d$,
  $n_p\assign f$, and $s\assign s-d$.
\end{enumerate}

\paragraph{Scaled Hypergeometric Approximation}

One possibility for modifying the values of $s$ and $f$ to let $s/(s+f) = q_t$ before applying $\Pr(n_t=m)=\binom{s}{m}\binom{f}{n_p-m}/\binom{s+f}{n_p}$, while maintaining nonzero and reasonable probabilities for all possible subset sizes, is to scale up linearly. This maintains at least some of the \emph{without replacement} property of the hypergeometric distribution, while the balance of left and right subrange is set to reflect statistical estimates. We have the following cases. 

If $q_t=0$, given that $q_t$ values are to be trusted, we know with certainty that $n_t=0$. If $q_t$ merely reflects statistics over some training data, a probability range should be reserved for this case, the size of which may need adjusting dependent of the application. In our tests, we simply assume that the $q_t$ values are correct (as they are in the experimental setting), and that $n_t$ does not have to be  explicitly represented.

If $q_t=1$, we have, analogously, that $n_t=\min\{n_p, s\}$.

If $s/f \ge q_t/(1-q_t)$, reassign $f\assign \round{s(1-q_t)/q_t}$.

If $s/f < q_t/(1-q_t)$, reassign $s\assign \round{fq_t/(1-q_t)}$.

The transform may appear as somewhat ad hoc, but experiments indicate good performance, in particular when prepended with the \emph{case exclusion} transform described above.

\paragraph{Non-Central Hypergeometric Distribution}

A less ad-hoc way of adjusting the hypergeometric distribution for the
statistical case is to employ \emph{Wallenius' non-central hypergeometric
  distribution}~\cite{FogCalcNchg}, henceforth referred to as \textsc{nchg}. \textsc{nchg}
is a generalization of the hypergeometric
distribution where a weight $w$ introduces a bias between success
and failure states. We set $w=f/s \times q_t/(1-q_t)$ for a suitable bias.

Computationally, \textsc{nchg} is considerably more challenging than the
previously listed distributions. We are aware of no closed form to produce the
desired probability ranges exactly. For the \emph{txt} data sets, we compute values of 1.14 and 1.62 bits per element, the most successful for the small sets, but we have been unable to compute the probabilities for the moderate-sized sets. Most likely, this is too
inefficient to be considered.

\medskip

\noindent Table~\ref{tab-stats} shows measurements for the binomial and rescaled hypergeometric distributions. As expected, the improvement from table~\ref{tab-nonstat} is significant.


\begin{table}[tbp]
\footnotesize
\begin{center}
\tabcolsep 2pt
\def\arraystretch{1.1}
\begin{tabular*}{.4\textwidth}{@{\extracolsep{\fill}}rlcccc}

&\multicolumn{1}{c}{} & \multicolumn{1}{c}{\textit{words}} & \multicolumn{1}{c}{\textit{inverted}} \\

\hline
{\scriptsize 1} &\textit{\rss binomial} & 3.54 & 3.26 \\
{\scriptsize 2} &\textit{\rss rescaled hg} & 3.48 & 3.22 \\
\end{tabular*}
\end{center}
\caption{\small\label{tab-stats}Encoding lengths (bits per element) of codes using element probabilities.}

\end{table}
\begin{table}[tbp]
\footnotesize
\begin{center}
\tabcolsep 2pt
\def\arraystretch{1.1}
\begin{tabular*}{.4\textwidth}{@{\extracolsep{\fill}}rlccc}

&\multicolumn{1}{c}{} & \multicolumn{1}{c}{\textit{A}} &\multicolumn{1}{c}{\textit{B}} &\multicolumn{1}{c}{\textit{C}} \\

\hline
{\scriptsize 1} &\textit{interpolative} & 8.63 & 5.96 & 1.43 \\
{\scriptsize 2} &\textit{tournament} & 7.08 & 12.17 & 0.91 \\
{\scriptsize 3} &\textit{\rss binomial} & 1.40 & 1.39 & 6.96\\
{\scriptsize 4} &\textit{\rss rescaled hg} & 1.39 & 1.37 & 8.26 \\
\end{tabular*}
\end{center}
\caption{\small\label{tab-extreme}Encoding lengths (bits per element) for generated extreme sets.}
\end{table}

\subsection{Extreme Element Distributions and Universe Permutation}\label{sec-uniperm}\label{sec-extreme}

We now study the behavior on some elaborate variations in input data. First, consider the extreme case where $S$ consists only of numbers divisible by $k$, and this is predicted correctly by the values of $q_t$. With $k=100$ and $|U|=10\,000$, we get the results in column~A of table~\ref{tab-extreme}.

This extreme can partially explain how \rss captures properties that range-narrowing codes do not: the encoding length is zero for the lower $\sim \log_2 k$ levels. To the range-narrowing codes elements are spread out evenly on each recursion depth. The behavior is somewhat similar even if the $|U|/k$ elements that are the only ones to appear are randomly distributed across $U$ (column~B). (Tournament coding degenerates when the maximum element is near the low end of the range.) On the other hand, if the $|U|/k$ elements lie only on one end of $U$s range, we have the most skewed distribution, and the most compressible case for all the recursive methods, shown in column~3.

Finally, with access to global element probabilities, a simple modification of encoding is to permute the enumeration of $U$ in probability order. As seen by comparing tables~\ref{tab-permut} and~\ref{tab-nonstat}, this has a strong effect on range-narrowing codes.  Also \rss benefits somewhat compared to table~\ref{tab-stats}, since skewness in the subset size distribution increases on higher levels of the tree.

\begin{table}[tbp]
\footnotesize
\begin{center}
\tabcolsep 2pt
\begin{tabular*}{.4\textwidth}{@{\extracolsep{\fill}}rlcc}

&
\multicolumn{1}{c}{\textit{words}} & \multicolumn{1}{c}{\textit{inverted}} \\

\hline
{\scriptsize 1} &\textit{interpolative} & 2.68 & 2.78 \\
{\scriptsize 2} &\textit{tournament} & 2.79 & 2.95 \\
{\scriptsize 3} &\textit{\rss binomial} & 2.92 & 2.72 \\
{\scriptsize 4} &\textit{\rss rescaled hg} & 2.97 & 2.81 \\
\end{tabular*}
\end{center}
\caption{\label{tab-permut}Encoding lengths (bits per element) with alphabet permuted in probability order.}
\end{table}


\section{Conclusion and Future Research}\label{sec-concl}

We have demonstrated several ways of exploiting statistical knowledge of elements when compressing integer sets, which has opened a number of paths for future research. The methods for which we have the strongest indication of good compression performance are still rather crude in some aspects. For instance, permutation based on global probability, explored in the final section above, still does not address the issue of \emph{context}, element probabilities conditioned on the presence of other elements. Consider, for example, the case where the global element probabilities are all approximately the same, but where there is a strong correlation leading certain elements scattered across the range of elements to frequently appear in the same sets. Permuting the range in probability order clearly does not capture such a regularity, and although the most simple cases would be easy to handle, using a general method to detect correlations, e.g., min-wise hashing~\cite{broderminhash}, for probabilistic modeling of set compression, is nontrivial, and an interesting topic for future research.

Other areas to explore include set compression on a wider application area than integer strings. Recursive subset-size encoding may be applicable also for bitstrings of non-homogeneous or unlimited lengths.

\small
\bibliography{jesper}

\end{document}